\documentclass[12pt]{article}

\def\SHOWAUTHORS{}
\usepackage[margin=1in]{geometry}
\usepackage{amsmath,amssymb}
\usepackage{booktabs}
\usepackage{graphicx}
\usepackage{microtype}
\usepackage{setspace}
\usepackage{placeins}
\usepackage{float}

\usepackage{mathtools}
\usepackage{natbib}
\usepackage[hidelinks]{hyperref}

\graphicspath{{fig/}}

\newcommand{\incfig}[2][]{%
  \IfFileExists{fig/#2}{\includegraphics[#1]{fig/#2}}%
  {\fbox{\footnotesize Missing file: \texttt{\detokenize{#2}}}}%
}

\DeclareMathOperator{\tr}{tr}
\DeclareMathOperator{\AM}{AM}
\DeclareMathOperator{\GM}{GM}

\newcommand{\sph}{S}
\newcommand{\svar}{S_{\mathrm{var}}}
\newcommand{\scorr}{S_{\mathrm{corr}}}

\newif\ifshowauthors
\showauthorsfalse
\ifdefined\SHOWAUTHORS\showauthorstrue\fi
\ifshowauthors
  \IfFileExists{authors.tex}{}{%
    \showauthorsfalse
    \GenericWarning{}{authors.tex not found: building the ANONYMOUS version}}%
\fi
\ifshowauthors
  \makeatletter
  \renewcommand*{\@fnsymbol}[1]{\ensuremath{\ifcase#1\or *\or \dagger\or
    \ddagger\or \mathsection\or \mathparagraph\else\@ctrerr\fi}}
  \makeatother
\fi

\title{Factoring A-Optimality into D-Optimality and Sphericity}
\hypersetup{
  pdftitle={Factoring A-Optimality into D-Optimality and Sphericity},
  pdfauthor={},
  pdfsubject={A- and D-optimal experimental design},
  pdfkeywords={coefficient-estimate correlation, Mauchly sphericity, I-optimality, screening experiments}
}
\ifshowauthors
\author{Andrew T. Karl\thanks{Corresponding author. Karl Statistical Services LLC, Aurora, CO, USA. Email: \texttt{akarl@asu.edu}. ORCID: 0000-0002-5933-8706.}}
\hypersetup{pdfauthor={Andrew T. Karl}}
\newcommand{\manuscriptacknowledgments}{The author thanks Bradley Jones for helpful discussions and suggestions during the development of this work. The author also thanks Valerii Fedorov for his suggestion, forwarded by Jones, to use dispersion profiles based on Equation~(2.1.10) of \citet{FedorovHackl2025}.}

\else
  \author{}
\fi
\date{}

\begin{document}
\maketitle
\ifshowauthors\else
  \vspace{-\baselineskip}
\fi

\begin{abstract}
\noindent
The D criterion measures the volume of the joint confidence ellipsoid for
the linear model coefficients and ignores its shape, so designs with the
same D value can estimate individual effects with different variances.
Meanwhile, A-optimality minimizes average coefficient variance. With both criteria
expressed as information values, A equals D multiplied by a sphericity
index for the same ellipsoid. In a fixed coefficient basis, sphericity
further factors into coefficient-variance balance and a determinant-based
correlation component. In five published screening comparisons, the
A-optimal design has a larger correlation component despite slightly
poorer variance balance; in three it also has a smaller D value.
In a seven-run family of designs that all tie under D, the two with equal
coefficient variances have the lowest A value. Both sphericity components can be calculated
directly from standard errors and estimate correlations available in
statistical software. After whitening by a prediction moment matrix, the same
determinant--sphericity factorization applies to the I-criterion.
\end{abstract}

\section{Introduction}\label{sec:intro}

In a screening experiment, the goal is to find a subset of factors that
deserve further investigation. An imprecise estimate may hide an active
effect, while a false lead can consume the budget for a follow-up experiment.
Strong correlations among coefficient estimates can also make active
effects harder to identify. D-optimality minimizes
the volume of the joint confidence ellipsoid for the model parameters,
whereas A-optimality minimizes the average variance of their estimates.
Both describe aspects of the precision of the parameter estimates, but they
can produce different designs with different properties.
\citet{JonesAllenMoyerGoos2020} compare the two criteria for small
screening experiments and argue for A-optimality as a default for
screening on these grounds. Their
examples include designs with equal D values and different coefficient
standard errors \citep[see also][]{StallrichAllenMoyerJones2023}.

The choice of criterion begins with the purpose of the experiment
\citep[pp.~56--57]{Fedorov1972}. The relative costs of missing an active
effect, pursuing an inactive one, and collecting more observations
determine which aspects of precision in parameter estimation deserve attention. Neither A nor D
incorporates all of these costs. As \citet[p.~277]{Kiefer1975b} emphasized,
a precisely stated criterion is often an approximation to a less precisely
formulated experimental objective. Geometric interpretations of A- and
D-optimality in terms of the joint confidence ellipsoid are presented by
\citet[Sec.~2.1]{FedorovHackl2025} and
\citet[Secs.~6.2--6.3]{Pukelsheim2006}.

With both criteria expressed as efficiencies relative to a common ideal
orthogonal benchmark (which need not be attainable at the chosen run
size), we show that A equals D multiplied by a sphericity index for the
joint confidence ellipsoid.
The index attains its maximum value of one exactly when the ellipsoid is a
sphere, and it is a power of Mauchly's statistic for covariance sphericity
\citep{Mauchly1940}. This factorization can help practitioners understand why the two criteria
sometimes favor different designs. A depends on the shape of the joint
confidence ellipsoid as well as its volume.

Following the decomposition in \citet[eq.~(12)]{Mauchly1940}, the
sphericity index further factors into two indices on the unit interval:
one for balance of the individual parameter estimate variances, and one for
coefficient-estimate correlation, based on the determinant of the
correlation matrix. With the coefficient basis and coding fixed, this
factorization quantifies how a design can give up some D value and still
improve A through a larger product of the two components. Both components can be computed from standard errors and estimate
correlations already displayed by software such as JMP
(Supplementary Section~S1).

Although the factorization also applies to other models, such as
response surface models, the main A--D comparisons focus on main-effects
screening models, with or without two-factor interactions, in fixed coded
units. They consist of the five published A--D pairs of
\citet{JonesAllenMoyerGoos2020} and an exact D-tie family based on
\citet{StallrichAllenMoyerJones2023}.

Finally, the same determinant--sphericity factorization extends to the
I criterion once the information matrix is whitened by the prediction
moment matrix (Section~\ref{sec:target}). The whitening step follows the I-to-A
transformation described by \citet[Sec.~5.3]{FilovaHarman2020}.

\section{The design problem}\label{sec:setting}

Suppose that $k$ experimental factors have been coded on
$\mathcal X=[-1,1]^k$. An exact $n$-run design chooses points
$x_1,\ldots,x_n\in\mathcal X$, with replication allowed. For a
specified regression vector $f(x)$ containing $p$ model terms, the
$i$th row of the model matrix $X$ is $f(x_i)^\top$. In a
main-effects model (ME),
\[
f(x)=(1,x_1,\ldots,x_k)^\top.
\]
The ME+2FI model adds each product $x_jx_\ell$ with $j<\ell$.
Thus $p=1+k$ or $p=1+k+\binom{k}{2}$, respectively.

Consider the full-rank linear model
\[
Y=X\beta+\varepsilon,\qquad
E(\varepsilon)=0,\qquad
\operatorname{Var}(\varepsilon)=\sigma^2I_n.
\]
Write $C=X^\top X$ for the information matrix and
$\Sigma=C^{-1}=\operatorname{Var}(\hat\beta)/\sigma^2$ for the scaled
covariance matrix of the coefficient estimates.
For fixed $n$, region, model, and coding, a D-optimal design minimizes
$\det(\Sigma)$ and an A-optimal design minimizes $\tr(\Sigma)$.
The common error variance $\sigma^2$ is omitted from here on.

\section{Factoring the A criterion}\label{sec:factorization}

\subsection{The determinant and covariance shape}\label{sec:sphericity}

For the factorization, the D and A criteria are expressed as information values:
\begin{equation}\label{eq:classical}
\Phi_0(C)=\det(C)^{1/p},\qquad
\Phi_{-1}(C)=\frac{p}{\tr(\Sigma)},
\end{equation}
where larger values are better. These are the geometric and harmonic matrix
means of $C$, following \citet[Secs.~6.1 and 6.7]{Pukelsheim2006}.
In this information-scale indexing, D and A are the order $0$ and $-1$
members of Kiefer's criterion family
(\citealp{Kiefer1974}; see also \citealp[Sec.~6.7]{Pukelsheim2006}).
Their reciprocals give equivalent dispersion criteria, for which smaller
values are better. Section~\ref{sec:profiles} plots the whole family on
this dispersion scale, showing how design comparisons change with the
order of the criterion.

Let $\lambda_1,\ldots,\lambda_p$ be the eigenvalues of
$\Sigma$, and denote arithmetic and geometric means by $\AM$ and
$\GM$. Then $\Phi_0=1/\GM(\lambda)$ and
$\Phi_{-1}=1/\AM(\lambda)$. Their ratio is
\begin{equation}\label{eq:stageone}
\sph_{-1}(C)=\frac{\Phi_{-1}(C)}{\Phi_0(C)}
=\frac{\det(\Sigma)^{1/p}}{\tr(\Sigma)/p}
=\frac{\GM(\lambda)}{\AM(\lambda)}.
\end{equation}
The arithmetic--geometric mean inequality gives
$0<\sph_{-1}\le1$, with equality precisely when
$\Sigma$ is proportional to the identity.
Equation~\eqref{eq:stageone} says that the A value is exactly the D value
multiplied by this sphericity factor.

This ratio describes the shape of the covariance ellipsoid: with normally
distributed errors and known error variance, the principal semi-axes of
the joint confidence ellipsoid are proportional to $\sqrt{\lambda_j}$. Equal eigenvalues
give a sphere; unequal eigenvalues reduce $\sph_{-1}$. Accordingly,
$\sph_{-1}$ is referred to here as \textit{sphericity}, or the
\textit{sphericity index}.

Incidentally, $\sph_{-1}$ is the $p$th root of Mauchly's $W$ statistic,
which is used for testing the sphericity of a covariance matrix in
multivariate analysis \citep{Mauchly1940}. Note that only the geometric interpretation of the index is
used here; Mauchly's distributional and hypothesis-testing framework plays
no role.

\subsection{Coefficient-variance balance and estimate correlations}\label{sec:afactor}

The sphericity index $\sph_{-1}$ of the confidence ellipsoid can be further
factored into two indices in $(0,1]$. One measures the balance of the
coefficient variances and the other measures the correlations among
the coefficient estimates.

Let
$d_j=\Sigma_{jj}$ be the variance of the $j$th coefficient estimate
apart from $\sigma^2$, put $V=\operatorname{diag}(d_1,\ldots,d_p)$,
and define
\begin{equation}\label{eq:rbeta}
R_{\hat\beta}=V^{-1/2}\Sigma V^{-1/2}.
\end{equation}
This is the correlation matrix of the coefficient estimates,
including the intercept. Since
$\Sigma=V^{1/2}R_{\hat\beta}V^{1/2}$,
\begin{equation}\label{eq:dettrace}
\det(\Sigma)=\left(\prod_{j=1}^p d_j\right)\det(R_{\hat\beta}),
\qquad
\tr(\Sigma)=\sum_{j=1}^p d_j.
\end{equation}
Let $\eta_1,\ldots,\eta_p$ be the eigenvalues of $R_{\hat\beta}$.
Since $R_{\hat\beta}$ has ones on its diagonal, its eigenvalues sum to $p$
and have arithmetic mean one. Their geometric mean is
$\det(R_{\hat\beta})^{1/p}$, so the correlation component can also be
expressed as the GM/AM ratio $\GM(\eta)/\AM(\eta)$.
Substituting these expressions into Equation~\eqref{eq:stageone} gives the
factorization of the A-criterion:
\begin{equation}\label{eq:nested}
\boxed{\begin{aligned}
\Phi_{-1}(C)&=\Phi_0(C)\,\sph_{-1}(C)\\
&=\Phi_0(C)
\frac{\GM(d)}{\AM(d)}
\det(R_{\hat\beta})^{1/p}\\
&=\Phi_0(C)
\frac{\GM(d)}{\AM(d)}
\frac{\GM(\eta)}{\AM(\eta)}.
\end{aligned}}
\end{equation}
Define the variance-balance component $\svar=\GM(d)/\AM(d)$ and the
correlation component $\scorr=\GM(\eta)/\AM(\eta)$, so that
$\sph_{-1}=\svar\scorr$. Their separate interpretations depend on
the coefficient basis and coding (Section~\ref{sec:coding}).
This decomposition mirrors that of \citet[eq.~(12)]{Mauchly1940}.

Each of the three sphericity indices is a ratio of a geometric mean
to an arithmetic mean, so none can exceed one. These bounds are
special cases of the trace--determinant inequality in
\citet[eq.~(B.3.5)]{FedorovHackl2025}.
The coefficient-variance component $\svar$ equals one exactly when
all $d_j$ are equal. The determinant-based correlation component
$\scorr$ equals one exactly when $R_{\hat\beta}=I$. With normally distributed
errors, this is also the condition for independence of the estimates.
For uncorrelated estimates, the coefficient axes are principal axes of the
covariance ellipsoid; equal coefficient variances then make it a sphere.
The index $\scorr$ summarizes the full correlation matrix, so a larger
value need not mean that every pairwise correlation is closer to zero.

Designs with the same coefficient variances have the same A value, even
if their correlations differ. Among such designs, a smaller determinant
of $R_{\hat\beta}$ reduces the confidence-ellipsoid volume and raises the
D value, while $\scorr$ decreases by the reciprocal factor. For designs with the same D value, a larger product
$\sph_{-1}=\svar\scorr$ gives a larger A value.

\subsection{Coding and the choice of effects}\label{sec:coding}

Equation~\eqref{eq:nested} holds for any positive-definite information
matrix, but its component values depend on the coefficient basis and
coding. Under a nonsingular change of basis $X^*=XT$, the information matrix
becomes $C^*=T^\top CT$ and $\Phi_0(C^*)=|\det T|^{2/p}\Phi_0(C)$, meaning
D rankings are unchanged when the same transformation is applied to every
design. However, A rankings can differ because recoding alters the
coefficient variances \citep[Secs.~6.2--6.3]{Pukelsheim2006}.
Separate rescalings of the model columns leave $\scorr$ unchanged
but generally change $\svar$ and the D value.

Orthogonal recoding preserves both criterion values and
$\sph_{-1}$, but can assign all departure from sphericity to
either component. For a given design, an eigenvector basis gives
$(\svar,\scorr)=(\sph_{-1},1)$. The Schur--Horn theorem also
guarantees a basis with equal coefficient variances, for which
$(\svar,\scorr)=(1,\sph_{-1})$
\citep[Ch.~9, B.2]{MarshallOlkinArnold2011}.
These rotations change the coefficient basis, but transforming the
estimates back to the named effects recovers their original covariance
matrix.

The screening comparisons follow
\citet{JonesAllenMoyerGoos2020} in coding each factor's chosen low
and high settings as $-1$ and $1$.
A unit change then corresponds to half of that factor's experimental
range. Changing the engineering units leaves the coded model matrix, and
hence the A value, unchanged as long as the physical low and high settings
remain the same.
Interpretations of the components of Equation~\eqref{eq:nested} refer to
the specified model terms in these specific coded units.

As in their full-model A and D comparisons, the intercept is retained in
both criteria and all components.
In situations where only a specified subset of coefficients is of interest, the
remaining coefficients can be treated as nuisance parameters. For example,
\citet[Sec.~2 and Secs.~3.1--3.2]{StallrichAllenMoyerJones2023}
treat the intercept as a nuisance parameter in their main-effects
screening comparisons.
Projecting the nuisance model columns out of the columns of interest gives
an information matrix equal to the corresponding Schur complement.
Its inverse is the covariance matrix of the estimates of interest,
apart from $\sigma^2$. When $p$ is replaced by the number of coefficients of interest, the same
factorization relates the information-value forms of the subset criteria
$A_s$ and $D_s$ used in that work, where the subscript $s$ indicates
restriction to those coefficients.

\section{Design comparisons}\label{sec:examples}

\subsection{Five published A--D comparisons}\label{sec:catalog}
\citet{JonesAllenMoyerGoos2020} present five examples comparing the
variance properties of A- and D-optimal screening designs: three for
ME models and two for ME+2FI models. The criteria and components below are recomputed from
their published factor-level matrices. Define the A and D
efficiencies relative to the ideal
orthogonal benchmark as
\begin{equation}\label{eq:scores}
a_n=100\frac{\Phi_{-1}(C)}{n},\qquad
d_n=100\frac{\Phi_0(C)}{n}.
\end{equation}
These scores equal 100 when $C=nI$, corresponding to the ideal
orthogonal benchmark in the specified units. The core factorization in
Equation~\eqref{eq:nested} may be written
\begin{equation}\label{eq:scoreid}
a_n=d_n\svar\scorr.
\end{equation}

Table~\ref{tab:catalog} shows the relative change from each
D-optimal design to its A-optimal partner. In all five comparisons,
the A-optimal design has a slightly smaller variance-balance component and a
larger correlation component. The first two pairs tie under D; in the
other three, D decreases. Since A increases in every pair, total
sphericity must increase by Equation~\eqref{eq:scoreid}. The larger correlation
component more than compensates for the small loss in variance balance.
The full scores and component values,
including the two orthogonal-array controls from the source, are
given in Supplementary Section~S3.

\begin{table}[!htbp]
\centering
\small
\begin{tabular}{lrrrr}
\toprule
Model, $(n,k)$ & $d_n$ & $\svar$ & $\scorr$ & $a_n$ \\
\midrule
ME, $(7,4)$       & $0.00$ & $-0.03$ & $5.30$ & $5.26$ \\
ME, $(15,9)$      & $0.00$ & $-0.01$ & $2.95$ & $2.94$ \\
ME, $(11,10)$     & $-0.46$ & $-0.04$ & $1.25$ & $0.75$ \\
ME+2FI, $(12,4)$  & $-2.58$ & $-0.83$ & $18.57$ & $14.55$ \\
ME+2FI, $(28,6)$  & $-1.83$ & $-0.30$ & $4.34$ & $2.12$ \\
\bottomrule
\end{tabular}
\caption{Percentage changes from the D-optimal to the A-optimal
design in each published pair. Rows are labeled by model and by
$(n,k)$, with $n$ runs and $k$ factors. Each entry is
$100\{Q(C_{\mathrm A})/Q(C_{\mathrm D})-1\}$ for the column quantity
$Q$. The $d_n$, $\svar$, and $\scorr$ ratios multiply to the $a_n$
ratio.}
\label{tab:catalog}
\end{table}

The 12-run comparison provides a particularly clean example. Its four main effects
and six two-factor interactions give eleven coefficients including
the intercept. The A scores for the A-optimal and D-optimal designs
are $80.00$ and $69.84$, respectively; their D scores are $83.56$
and $85.78$. Equation~\eqref{eq:scoreid} resolves
the A ratio as
\[
\frac{a_n(C_{\mathrm A})}{a_n(C_{\mathrm D})}
=\frac{d_n(C_{\mathrm A})}{d_n(C_{\mathrm D})}
 \frac{\svar(C_{\mathrm A})}{\svar(C_{\mathrm D})}
 \frac{\scorr(C_{\mathrm A})}{\scorr(C_{\mathrm D})}
\approx0.9742\times0.9917\times1.1857
\approx1.1455.
\]
Thus the A-optimal design gives up about $2.6\%$ in the D value and
$0.8\%$ in variance balance, but gains $18.6\%$ in the correlation
component. Its A value is $14.5\%$ larger, or equivalently its
average coefficient variance is about $12.7\%$ smaller.

\begin{figure}[!htbp]
\centering
\incfig[width=\linewidth]{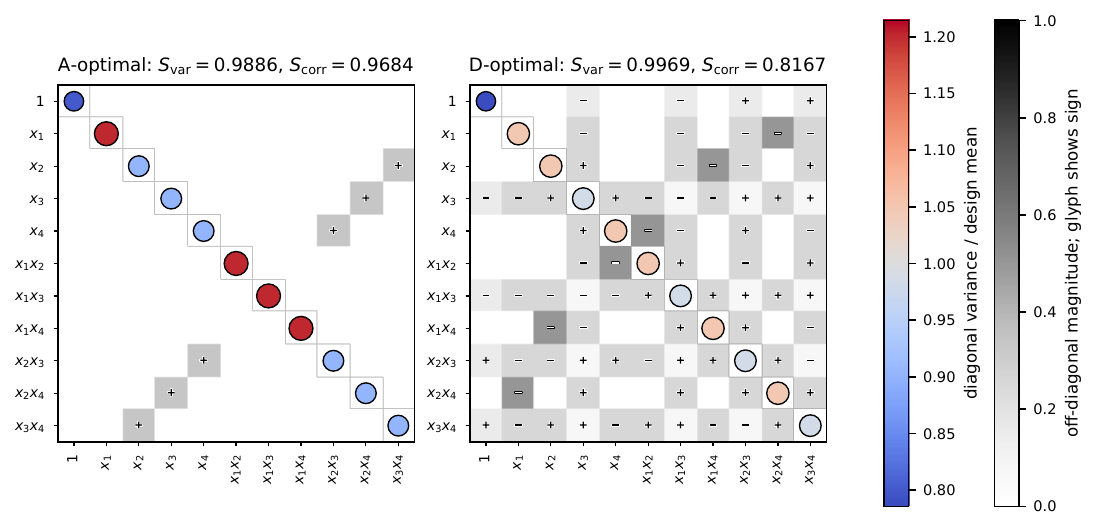}
\caption{Coefficient-estimate variances and correlations for the 12-run ME+2FI
A-optimal (left) and D-optimal (right) designs. All eleven coefficients,
including the intercept, are shown. Off-diagonal shade gives absolute
correlation on a common scale, with signs indicated by glyphs.
On the diagonal, circle area is proportional to
$d_j/\operatorname{AM}(d)$; the legend marks the reference value one. Diagonal sizes and colors compare variances relative to each
design's own mean. The mean coefficient variances are approximately
$0.1042\sigma^2$ for A and $0.1193\sigma^2$ for D.}
\label{fig:estimatecorr}
\end{figure}

Figure~\ref{fig:estimatecorr} shows the relative coefficient variances and
estimate correlations summarized by the two sphericity components in
Equation~\eqref{eq:nested}. Diagonal circle areas and colors show each coefficient variance relative
to the mean. Off-diagonal shade shows the absolute
pairwise correlations of the estimates, with signs indicated by
glyphs. Supplementary
Section~S1 shows how to calculate total sphericity from JMP's A and D
efficiencies and its two components from Design Std Errors and Correlation
of Estimates, including the required coding settings. It also compares the
coefficient-estimate map with the model-column correlation map. The two
displays describe different matrices: model-column correlations are
computed directly from $X$, while coefficient-estimate correlations are
obtained after inverting $X^\top X$.

In the plot, only 3 of the 55 estimate pairs are
correlated in the A-optimal design, compared with 37 in the D-optimal design.
The A-optimal design has slightly less balanced variances but a larger
correlation determinant. The map identifies which coefficients are
involved; $\scorr$ summarizes their joint correlation pattern through
the determinant of the full correlation matrix.

\FloatBarrier

\subsection{An exact family tied under D}\label{sec:stallrich}

\citet[Fig.~1 and Cor.~1]{StallrichAllenMoyerJones2023} describe a seven-run,
five-factor design that is both A- and D-optimal for the main effects,
treating the intercept as a nuisance parameter. They
show that either of two entries in its first run can change without
altering the determinant. Let those entries vary jointly as
$(a,b)\in[-1,1]^2$, with the other runs fixed. The full factor-level
matrix and block calculation appear in Supplementary Section~S2.
For the information matrix including the intercept, the calculation gives
\begin{equation}\label{eq:stalldet}
\det\{C(a,b)\}=2^{16},\qquad
\tr\{C(a,b)^{-1}\}=1+\frac{q(a,b)}{64},\qquad
q(a,b)=3a^2+2ab+3b^2.
\end{equation}
The determinant is constant throughout the square, so both coordinates
can vary jointly while the design remains D-optimal. Since
$q=2a^2+(a+b)^2+2b^2$, it is uniquely minimized at $(0,0)$ and
maximized at the two same-sign corners. The center has the
largest A value, and those corners have the smallest.

\begin{figure}[!htbp]
\centering
\incfig[width=\linewidth]{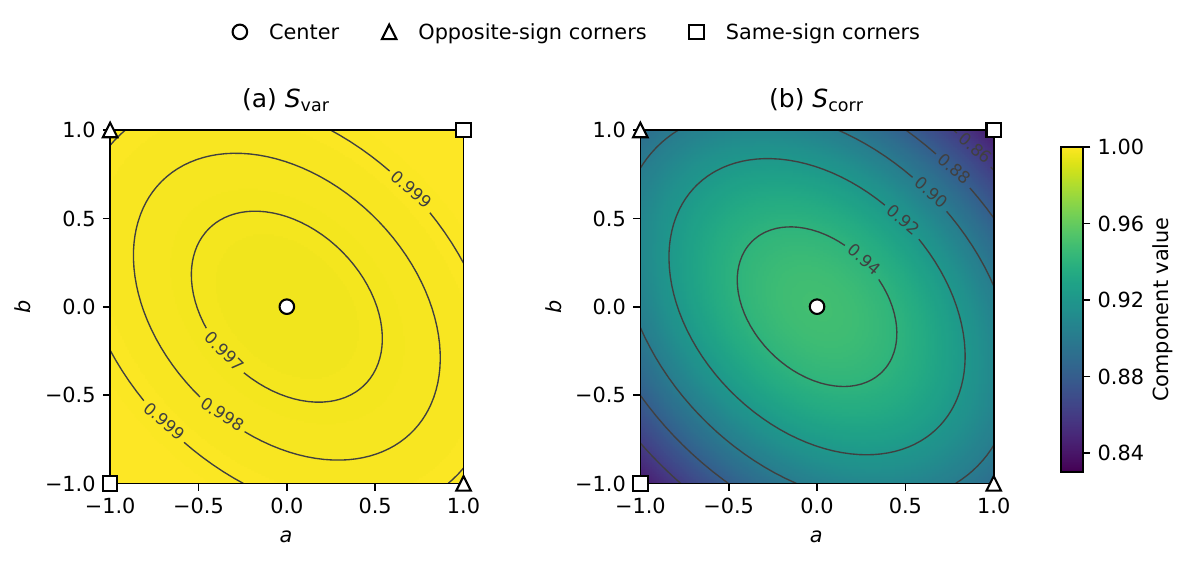}
\caption{Coefficient-variance sphericity and correlation sphericity
in the exact D-tie family. Both panels use the same color scale;
labeled contours give the component values. Symbols identify the
center, opposite-sign corners and same-sign corners. The product
of the two components determines the A ordering because D is constant.}
\label{fig:stallrichcomponents}
\end{figure}

The coefficient variances explain why the center is better than the
corners. In the order
$(1,x_1,\ldots,x_5)$ they are
\begin{equation}\label{eq:stalldiag}
\operatorname{diag}\{C(a,b)^{-1}\}
=\left(h,h,h,h,\frac{3}{16},\frac{3}{16}\right),
\qquad h=\frac{40+q}{256}.
\end{equation}
At a same-sign corner, $q=8$ and all six variances equal $3/16$.
At the center, four fall to $5/32$ and the other two remain
$3/16$. The center is better under A because four variances are smaller and the
other two are unchanged, even though they are now less balanced. Perfect balance at the
same-sign corners comes with a lower $\scorr$, as shown in
Table~\ref{tab:stallrich}. Throughout this family, $\svar$ ranges only from
$0.9962$ to one, while $\scorr$ ranges from $0.8399$ to $0.9485$.
Figure~\ref{fig:stallrichcomponents} shows the contrast on a common
scale. As $q$ decreases toward zero, the increase in $\scorr$
offsets the decrease in $\svar$, so their product---and hence A
at the common D value---increases.

\begin{table}[!htbp]
\centering
\small
\begin{tabular}{lrrrrr}
\toprule
Design & $a_n$ & $d_n$ & $\svar$ & $\scorr$ & $\sph_{-1}$ \\
\midrule
Center $(0,0)$       & 85.71 & 90.71 & 0.9962 & 0.9485 & 0.9449 \\
Corner $(1,1)$       & 76.19 & 90.71 & 1.0000 & 0.8399 & 0.8399 \\
Corner $(-1,1)$      & 80.67 & 90.71 & 0.9992 & 0.8901 & 0.8894 \\
\bottomrule
\end{tabular}
\caption{Three members of the exact D-tie family. The criterion
scores use the common ideal orthogonal benchmark; larger is better.
The same-sign corners have equal coefficient variances and the lowest
A value in the family.}
\label{tab:stallrich}
\end{table}

\FloatBarrier

\subsection{Dispersion profiles}\label{sec:profiles}

The same designs can be compared across the nonpositive information orders of
Kiefer's criterion family \citep{Kiefer1974}. The information values extend
Equation~\eqref{eq:classical}, and their reciprocals give the dispersion
profile of \citet[eqs.~(2.1.8) and (2.1.10)]{FedorovHackl2025}:
\begin{equation}\label{eq:dispersionprofile}
\begin{aligned}
\Phi_{-\gamma}(C)&=\{p^{-1}\tr(C^{-\gamma})\}^{-1/\gamma},
\qquad \gamma>0,\\
\Psi_\gamma(C)&=\frac{1}{\Phi_{-\gamma}(C)}
=\{p^{-1}\tr(\Sigma^\gamma)\}^{1/\gamma},\\
\Psi_0(C)&=\det(\Sigma)^{1/p}
\le\Psi_\gamma(C)\le\lambda_{\max}(\Sigma)=\Psi_\infty(C),
\end{aligned}
\end{equation}
where the subscripts zero and infinity denote limits. For the
dispersion profile, smaller values are better. The left limit is
$1/\Phi_0$ (the inverse D value), the value at $\gamma=1$ is
$1/\Phi_{-1}$ (the inverse A value), and the
right limit is the E criterion expressed as the largest covariance
eigenvalue. The profile is nondecreasing in $\gamma$
\citep[eq.~(B.3.8)]{FedorovHackl2025}.
This is the covariance formulation of the matrix means described by
\citet[Secs.~6.1 and 6.7]{Pukelsheim2006}: the order-$\gamma$ mean
of $\Sigma$ is the reciprocal of the order-$-\gamma$ mean of $C$.

\begin{figure}[!htbp]
\centering
\incfig[width=\linewidth]{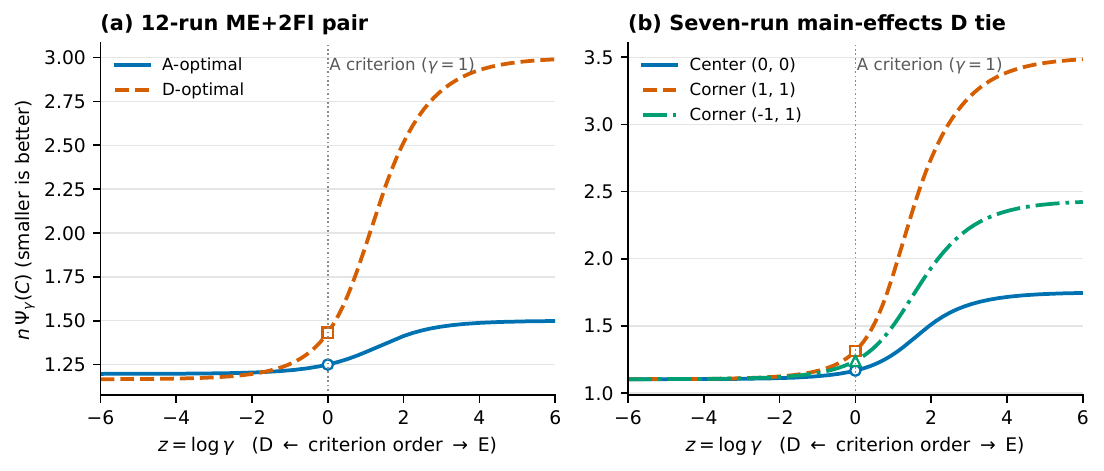}
\caption{Run-normalized dispersion profiles for (a) the published
12-run ME+2FI pair and (b) three members of the exact seven-run
D-tie family. Line styles distinguish designs; symbols mark the
A criterion at $\log\gamma=0$. The curves approach the D and E dispersion limits toward the left and
right, respectively.
Smaller values are better.}
\label{fig:profiles}
\end{figure}

Figure~\ref{fig:profiles} plots $n\Psi_\gamma(C)$ against
$\log\gamma$. For the 12-run pair, the D-optimal design begins
lower on the left of the plot, but the curves cross near
$\gamma=0.1805$; at $\gamma=1$ the A-optimal design is lower.
The three seven-run designs share their left-hand limit and separate
toward A and E, with the center $(0,0)$ design lowest at A and E.
These are the comparisons in Sections~\ref{sec:catalog} and
\ref{sec:stallrich}, now seen over Kiefer's family of criteria.

Writing
$\sph_{-\gamma}(C)=\Psi_0(C)/\Psi_\gamma(C)=\Phi_{-\gamma}(C)/\Phi_0(C)$
generalizes the sphericity index in Equation~\eqref{eq:stageone} from the
A-criterion to the nonpositive orders of Kiefer's criterion family.
Hence $\Phi_{-\gamma}(C)=\Phi_0(C)\sph_{-\gamma}(C)$, with the A--D
factorization at $\gamma=1$.
Different orders place different emphasis on departures from spherical
geometry. In the dispersion profiles, the ratio of each curve's height
to its left-hand limit is $1/\sph_{-\gamma}$; at
$\gamma=1$ it is the reciprocal of $\svar\scorr$.
The curve is flat exactly when the covariance is spherical.
A low profile with little relative rise from its left-hand limit indicates
small, similar variances along the principal axes of the confidence ellipsoid.
Moving right along the $\log\gamma$ axis places increasing emphasis on the
larger eigenvalues of $\Sigma$, or equivalently the smaller eigenvalues
of $C$. The E limit is determined by the variance in the least precisely estimated
direction, which is an eigenvector direction of $\Sigma$ and need not
coincide with a named coefficient axis. The plots thus show covariance shape together with
the precision summarized by the D value.
\FloatBarrier

\subsection{Factoring the I criterion}\label{sec:target}

In their screening comparisons, \citet[Sec.~5]{JonesAllenMoyerGoos2020}
report that A-optimal designs often have higher I-efficiency than
D-optimal designs. The I criterion measures average prediction variance over the factor
space. Let $\nu$ be a probability measure specifying the prediction
weights, common to all designs, with positive-definite prediction moment
matrix $B$ \citep[Sec.~2.2.3, Eqs.~(2)--(3)]{JonesAllenMoyerGoos2020}.
Using its Cholesky factor $L$, write
\begin{equation}\label{eq:targetB}
B=\int f(x)f(x)^\top\,d\nu(x)=LL^\top.
\end{equation}
The whitened information matrix is $C_B=L^{-1}CL^{-\top}$.
The first line of Equation~\eqref{eq:nested} relates A and D when
applied to $C$, and I and D when applied to $C_B$.
Supplementary Section~S4 explains how to obtain and normalize this moment
matrix from JMP's saved \texttt{Moments Matrix} output.

At a prediction point $x$, the fitted mean $f(x)^\top\hat\beta$ has
variance $\sigma^2 f(x)^\top C^{-1}f(x)$
\citep[Sec.~2.1]{FedorovHackl2025}. Define $\Sigma_B=C_B^{-1}$. Averaging this variance with respect to $\nu$
and dividing by $\sigma^2$ gives
\begin{equation}\label{eq:targettrace}
I_\nu(C)=\int f(x)^\top C^{-1}f(x)\,d\nu(x)
=\tr(C^{-1}B)=\tr(\Sigma_B).
\end{equation}
The trace equalities follow from $f^\top C^{-1}f=\tr(C^{-1}ff^\top)$,
$\Sigma_B=L^\top C^{-1}L$, and the cyclic property of the trace.
This identity expresses average prediction variance in the original
model as the sum of coefficient variances in the transformed model.
\citet[Sec.~5.3]{FilovaHarman2020} use this equivalence to compute
I-optimal designs with methods developed for A-optimality.
Define the larger-is-better
information value $\Phi_{-1}^B(C)=p/I_\nu(C)$. Applying
Equation~\eqref{eq:stageone} to $C_B$ gives
\begin{equation}\label{eq:targetIfactor}
\begin{aligned}
\Phi_{-1}^B(C)&=\Phi_0^B(C)\,\sph_{-1}(C_B),\\
\Phi_0^B(C)&=\frac{\Phi_0(C)}{\det(B)^{1/p}}.
\end{aligned}
\end{equation}
Maximizing $\Phi_{-1}^B(C)$ gives an I-optimal design.
Maximizing $\Phi_0^B(C)$ gives a D-optimal design, since
$\det(B)^{1/p}$ does not depend on the design.
The target-relative sphericity $\sph_{-1}(C_B)$ measures the shape of the
coefficient covariance in the whitened basis. It is at most one, with
equality exactly when $C$ is proportional to $B$. For fixed $B$, both
information values and target-relative sphericity are unchanged
by the choice of factor $L$ satisfying $B=LL^\top$ (Supplementary
Section~S4.1).
In the whitened basis, the variance-balance component $\svar$ and
correlation component $\scorr$ can change with the choice of $L$.
Only their product is therefore reported.

For the ME and ME+2FI models, \citet[Sec.~2.2.3]{JonesAllenMoyerGoos2020}
give diagonal moment matrices under uniform weighting on the coded
cube. For the unrescaled square-free terms (products of distinct factors)
used in these models, equal prediction weights on the $2^k$ cube vertices
give $B=I$, regardless of where the experimental runs are placed. Thus
$\tr(C^{-1})$ is the vertex-average prediction variance, and I- and
A-optimality are equivalent under those weights. In this coding,
$\sph_{-1}(C)$ is also the target-relative sphericity for the vertex
prediction target. Under a common nonsingular recoding of $C$ and $B$,
this value is unchanged.

The profile definition in Equation~\eqref{eq:dispersionprofile} also applies
to $C_B$. At $\log\gamma=0$ ($\gamma=1$), the run-normalized ordinate
is $nI_\nu(C)/p$, so this point represents the I criterion.
The left-hand limit still represents the D criterion, with its dispersion
rescaled by the common factor $\det(B)^{1/p}$.

\begin{figure}[!htbp]
\centering
\incfig[width=\linewidth]{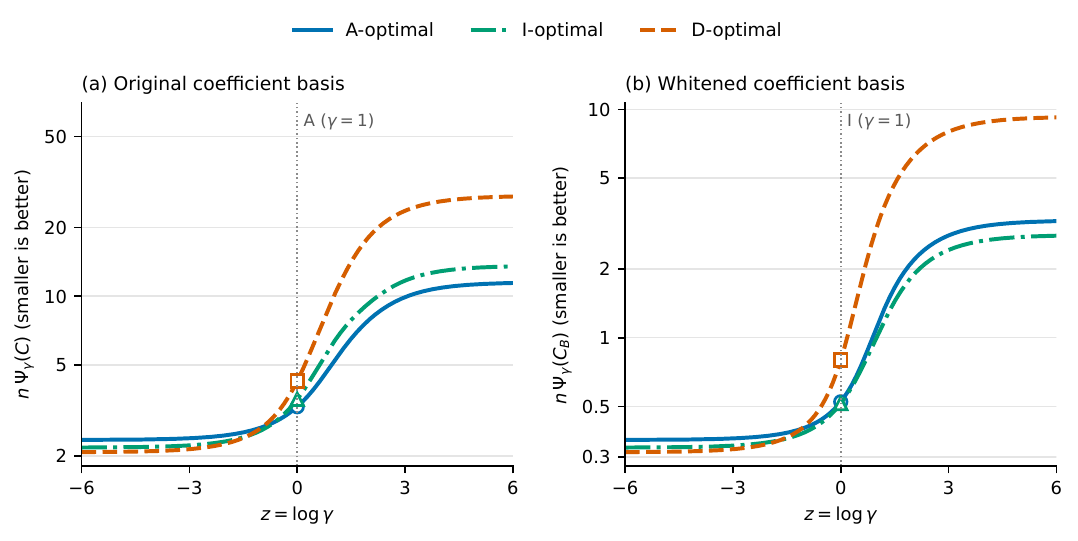}
\caption{Run-normalized dispersion profiles for the three supplied
26-run full-quadratic designs: (a) the original coefficient basis and
(b) the basis whitened for uniform prediction weights on $[-1,1]^5$.
Both ordinates are logarithmic, with the same scale per logarithmic
unit. Symbols at $\log\gamma=0$ mark A in (a) and I in (b).
The left-hand limits are the respective D dispersions; the
right-hand limits are $n\lambda_{\max}(C^{-1})$ and
$n\lambda_{\max}(C_B^{-1})$. The ratio of the ordinate at
$\gamma=1$ to its left-hand limit is reciprocal total sphericity.
Smaller dispersions are better.}
\label{fig:targetprofiles}
\end{figure}

Figure~\ref{fig:targetprofiles} compares three 26-run designs for a
five-factor full quadratic model ($p=21$), with prediction averaged
uniformly over the continuous cube $[-1,1]^5$. The A, I, and D designs
were generated with the corresponding criteria in JMP Custom Design
(Supplementary Section~S4). The original basis
uses the unrescaled main effects, interactions, and uncentered pure
quadratic terms. The two panels show the profiles before and after
whitening. Their D rankings are the same, but the A design is lowest
at $\gamma=1$ in the original basis and the I design is lowest there
in the whitened basis.

For the D and I designs, the factorization gives the ratio of I
efficiencies
\[
\frac{\Phi_{-1}^B(C_{\mathrm D})}{\Phi_{-1}^B(C_{\mathrm I})}
=\frac{\Phi_0^B(C_{\mathrm D})}{\Phi_0^B(C_{\mathrm I})}
 \frac{\sph_{-1}(C_{B,\mathrm D})}{\sph_{-1}(C_{B,\mathrm I})}
\approx 1.048\times0.616\approx0.646.
\]
The D design's larger transformed D value is outweighed by its
smaller target-relative sphericity. The I design has about $35.4\%$ lower
average prediction variance. For the A and I designs, the ratio of
D values is $0.9257$ in both bases. The A design's sphericity is
$16.5\%$ larger in the original basis, sufficient to give it the
larger A value, but only $6.2\%$ larger after whitening. Its I
dispersion is $0.5240$, compared with $0.5153$ for the I design:
the I design predicts slightly better despite lower target-relative
sphericity. As discussed in Section~\ref{sec:coding}, the A comparison
in the original basis depends on how the quadratic model is coded.
For a fixed prediction target, the I comparison and target-relative
sphericity are unchanged by a common nonsingular recoding of $C$ and $B$.
Supplementary Section~S4 gives further details including invariance proofs
and the full numerical comparison in Table~S2.

\FloatBarrier

\section{Discussion}\label{sec:discussion}

Because $a_n=d_n\sph_{-1}=d_n\svar\scorr$, a gain in the sphericity
of the confidence ellipsoid can offset a loss in D between two competing
designs and give a larger A value. In the published
pairs, the gain in the correlation component outweighs a small loss in
variance balance and, in three cases, a loss in D. In the exact D-tie family, the designs with equal coefficient variances
have the lowest (worst) A value.

For teaching, the factorization connects average coefficient variance and
ellipsoid volume through covariance shape. At a fixed D value, greater
sphericity gives a larger A value. After whitening by the prediction
moment matrix, the same relationship holds for the I information value:
the D value is rescaled by a constant common to all designs, and
target-relative sphericity supplies the shape factor. In the response surface example,
the A design's sphericity advantage over the I design is smaller after
whitening, and the two designs reverse order under the A and I criteria.
The dispersion profiles display these comparisons across the Kiefer
criterion family.

The components $\svar$ and $\scorr$ describe variance balance and the
correlation structure of the estimates. They can be calculated from standard JMP output
(Supplementary Section~S1) and reported alongside the A and D values. The
coefficient-estimate map identifies the individual variances and
correlations involved. The choice among designs also depends on the costs
of additional runs and mistaken screening decisions; the components do
not determine those costs. When particular effects deserve greater weight,
their variances can be examined directly or incorporated in a weighted
criterion \citep{AllenMoyerStallrich2022}.

Neither sphericity nor either component should be optimized in isolation.
A spherical covariance can still have large variances, and the best
feasible design need not attain sphericity one. The allocation between
variance balance and correlation also depends on the basis in which the
effects are named. Within a specified model, coding, and prediction target,
the factorizations help interpret the differences among A, D, and I.

\section*{Funding}
No funding was received for this work.

\section*{Declaration of Generative AI Use}
OpenAI's GPT-5.6 Pro, GPT-6 Astra and Codex and Anthropic's Claude
Fable 5.1 were used under the author's direction to facilitate
exploration, implementation, and presentation. Uses included
brainstorming, adversarial critique, drafting, restructuring, rewording,
proofreading, and LaTeX formatting. The systems also assisted in
developing and documenting Python and JSL code and R design
transcriptions for verification, figures, examples, and design
reconstruction. Other uses were crosschecking citations and attributed
claims against a local PDF library and reviewing reproducibility and
packages. Claude Fable 5.1 initially derived the determinant and
block-inverse calculations reported in Supplementary Section~S2.
The research questions and central ideas, including the connection
between A/D/I and Mauchly's $W$ and the coefficient-diagnostic correlation
map, originated with the author. The author evaluated, rejected, or
revised the tools' suggestions and checked the substantive mathematics,
sources, code, and numerical results. The author accepts full responsibility
for the manuscript's content.

\section*{Data Availability and Reproducibility}
The technical supplement is included with this arXiv version as an
ancillary PDF. A separate reproduction package contains the design
matrices analyzed here, the data underlying the figures and tables,
and code that reproduces the reported calculations and graphics. Its
README documents the inputs and software environment. The reproduction
package is not included in this arXiv deposit.

\ifshowauthors
  \ifdefined\manuscriptacknowledgments
    \section*{Acknowledgments}
    \manuscriptacknowledgments
  \fi
\fi

\section*{Disclosure Statement}
The author reports no competing interests.

\bibliographystyle{apalike}
\bibliography{A_note}

\end{document}